\documentclass[useAMS,usegraphicx]{mn2e}

\title[Neural network identification of new $z \ge 3.6$ QSOs]{Use of neural networks for the identification of new $z \ge 3.6$ QSOs from FIRST-SDSS DR5}
\author[R. Carballo et al]
{R. Carballo,$^1$
\thanks{E-mail:carballor@unican.es}
  J.I. Gonz\'alez-Serrano,$^2$
  C.R. Benn,$^3$
  F. Jim\'enez-Luj\'an$^{2,4}$\\
 $^1$ Dpto. de Matem\'atica Aplicada y Ciencias de la Computaci\'on, Univ. de  Cantabria\\
       ETS Ingenieros de Caminos, Canales y Puertos, Avda de los Castros s/n,
 E-39005 Santander, Spain\\
        $^2$ Instituto de F\'\i sica de Cantabria (CSIC-Universidad de
Cantabria),  Avda de los Castros s/n,
 E-39005 Santander, Spain\\
$^3$Isaac Newton Group, Apartado 321, E-38700 Santa Cruz de La
Palma, Spain\\
$^4$ Dpto. de F\'\i sica Moderna,  Univ. de  Cantabria, Avda de los
Castros s/n,  E-39005 Santander, Spain\\}
\begin{document}

\date{Accepted .... Received ... ; in original form ...}

\pagerange{\pageref{firstpage}--\pageref{lastpage}} \pubyear{...}

\maketitle

\label{firstpage}

\begin{abstract}
We aim to obtain a complete sample of redshift $z \ge 3.6$ radio
QSOs from FIRST sources ($S_{\rm 1.4 ~GHz} > 1$ mJy) having
star-like counterparts in the SDSS DR5 photometric survey ($r_{\rm
AB} \le 20.2$). Our starting sample of 8665 FIRST-DR5 pairs includes
4250 objects with spectra in DR5, 52 of these being $z \ge 3.6$
QSOs.  We found that simple supervised neural networks, trained on
the sources with DR5 spectra, and using optical photometry and radio
data, are very effective for identifying high-$z$ QSOs in a sample
{\it without} spectra. For the sources with DR5 spectra the
technique yields a completeness (fraction of actual high-$z$ QSOs
classified as such by the neural network) of 96 per cent, and an
efficiency (fraction of objects selected by the neural network as
high-$z$ QSOs that actually are high-$z$ QSOs) of 62 per cent.
Applying the trained networks to the 4415 sources {\it without} DR5
spectra we found 58 $z \ge 3.6$ QSO candidates. We obtained spectra
of 27 of them, and 17 are confirmed as high-$z$ QSOs.  Spectra of 13
additional candidates from the literature and from SDSS DR6 revealed
seven more $z \ge 3.6$ QSOs, giving an overall efficiency of 60 per
cent (24/40).  None of the non-candidates with spectra from NED or
DR6 is a $z \ge 3.6$ QSO, consistently with a high completeness. The
initial sample of high-$z$ QSOs is increased from 52 to 76 sources,
i.e. by a factor 1.46. From the new identifications and candidates
we estimate an incompleteness of SDSS for the spectroscopic
classification of FIRST $3.6 \le z \le 4.6$ QSOs of 15 per cent for
$r \le 20.2$.
\end{abstract}


\section{Introduction}

Homogeneous statistical samples of high-redshift quasi-stellar objects
(QSOs) allow not only investigation of the QSO phenomenon itself, but
also provide important information for a wide variety of studies.
In particular, the luminosity
function of high-redshift QSOs provides strong constraints on the
theory of the accretion of matter onto supermassive black holes in the
nuclei of galaxies. The increasing evidence for a relation between
the formation of galaxy bulges and supermassive black holes (Kormendy
\& Richstone 1995, Magorrian et al. 1998) emphasises the importance of
understanding the role of QSO activity in the formation and evolution
of galaxies. The luminosity function of QSOs is also essential to
quantify their contribution to the X-ray background and the UV
ionising flux at high redshift.  In addition, the absorption spectra
of these QSOs reveal the state of the intergalactic medium at early
epochs.

Although radio-loud (RL) QSOs are a small subset of the QSO
population, samples of high-redshift RL QSOs benefit from higher
completeness, due to the drastically reduced contamination by stars in
samples of radio selected QSO candidates, compared to optically
(colour) selected QSO candidates (Richards et al. 2006).  Moreover,
the connection between radio and optical activity, which still needs
to be understood, requires a comparison between radio-loud and
radio-quiet QSO populations. Ivezik et al. (2004) provide conclusive evidence
that the distribution of radio-to-optical flux ratio for QSOs,
i.e. the radio loudness, is bimodal (the so-called QSO radio dichotomy),
on the basis of accurate optical and radio measurements of a large sample
of RL QSOs obtained from the Sloan Digital Sky Survey
(SDSS; York et al. 2000) and the
Faint Images of the Radio Sky at Twenty cm survey
(FIRST; Becker, White \& Helfand 1995).
Many studies suggest that RL QSOs reside in more massive
galaxies and harbour more massive central black holes than radio-quiet QSOs,
but the point is still controversial (see references for and against these
arguments at Cirasuolo et al. 2006).
A recent study by Jiang et al. (2007) based on a QSO sample drawn from SDSS
and FIRST shows that the fraction of RL QSOs decreases with increasing
redshift and with decreasing optical luminosity.

We aim to obtain a homogeneous sample of high-redshift RL
QSOs ($z$ above $\sim 3.6$) drawn from correlation of the FIRST
catalogue
%(Faint Images of the Radio Sky at 20 cm, Becker, White \& Helfand 1995,
($S_{\rm 1.4 ~GHz} > 1.0$ mJy) with unresolved
objects in the SDSS Data Release 5 (SDSS DR5, Adelman-McCarthy et al. 2007).
%The area covered by the DR5 imaging survey is $\sim$ 7893
The area of overlap between FIRST and the DR5 imaging survey is $\sim$ 7391
deg$^2$ and the number of selected FIRST-SDSS matches is 8665 (Section 2).
SDSS provides:
(i) $ugriz$ photometry, which is a powerful tool for separating
high-$z$ QSOs from other populations (e.g. stars, QSOs with
$z$ below $\sim 3.6$ or unresolved low-$z$ galaxies);
(ii) morphological classification, essential for distinguishing between high-$z$ QSOs
and galaxies or resolved low-$z$ active galactic nuclei; and
(iii) spectroscopy of many of our candidates (4250),
selected as spectroscopic targets by SDSS DR5.
Since SDSS spectroscopic observations necessarily lag the imaging,
the total DR5 spectroscopic area is lower,
% only dr5 $\sim$ 5740 deg$^2$
with $\sim 5553$ deg$^2$  included in the overlap with FIRST.
Most of the candidates with available spectroscopy were classified by
SDSS as QSOs, i.e. have a secure detection of a high-excitation
emission line with FWHM $\ge$ 1000 km sec$^{-1}$. The rest are
galaxies, stars and objects of `unknown' class. 52 DR5 sources were
spectroscopically classified as $z \ge 3.6$ QSOs (Table 1).

Our approach to obtaining a high-$z$ QSO sample was to extend the
existing sample of 52 FIRST-DR5  high-$z$ QSOs by applying automated
learning techniques, specifically neural networks (NNs), to the
8665 FIRST-SDSS DR5 photometric matches. NNs have been shown to
be powerful tools for both classification and regression
tasks, in many fields of astronomy, and have subsequently been applied to
predict object classes and/or astrophysical parameters.
Fields where NNs have been applied include:
 classification of stellar spectra (Bailer-Jones, Irwin \&
von Hippel 1998); morphological star/galaxy separation (e. g. Bertin
\& Arnouts 1996); morphological classification, spectral typing and/or
photometric redshifts of galaxies [Folkes, Lahav \& Maddox 1996;
Lahav et al. 1996; Firth, Lahav \& Somerville 2003; Collister \& Lahav
2004 (this paper reports the popular photometric redshift code ANNz);
Ball et al. 2004]; QSO identification and/or QSO photometric redshifts
(Carballo et al. 2004; Claeskens et al. 2006); and cross-matching of
astronomical catalogues (Rohde et al. 2005).

QSO selection and estimation of QSO photometric redshifts are of
prime importance for the SDSS project. Various studies address the
problem using different machine learning approaches. Richards et
al. (2004) applied a probability density analysis based on kernel
density estimation of the colour distribution of stars and
spectroscopically confirmed QSOs in SDSS DR1, to classify, as stars or QSOs,
a catalogue of over $10^5$ unresolved, $g \le 21$ mag, UV-excess
($u - g \le 1$) QSO candidates.
The resulting efficiency and completeness (the latter
evaluated for $g \le 19.5$) for the selection of QSOs in the candidate
sample was estimated to be around 95 per cent up to $z \simeq 2.4-3.0$,
the redshift limit mainly arising from the restriction of the catalogue to
UV-excess objects.

Suchkov, Hanish and Margon (2005) applied the oblique decision tree
classifier ClassX to classify SDSS-DR2 photometric objects into 25
classes [stars, red stars, 10 redshift bins for galaxies, and 13 for
Active Galactic Nuclei (hereafter AGN)] using colour information
and morphology (attributes `resolved' or `unresolved') from SDSS.
For each of the 12 redshift bins for AGN with $\Delta z = 0.2$
and covering $0 \le z \le 2.4$,
the completeness obtained for the test sample is in the range
from 43 to 81 per cent, with an average 63 per cent.
For the high-redshift bin, in the range $z=2.4-6.0$,
the completeness drops to 14 per cent,
the remaining high-$z$ AGNs beeing classifified as stars (47 per cent)
or as AGN in the adyacent redshift bin $2.2 \le z \le 2.4$ (39 per cent).
This result illustrates the difficulty in separating high-$z$ QSOs from
other classes. The efficiency or fraction of true high-$z$ QSOs
among the sources classified in the AGN high-$z$ bin was $\sim$ 75 per cent.

% l = [70.2 81.4 58.8 62.6 44.4 70.2 53.2 79.6 61.3 42.6 65.2 65.9]
% mean(l)  62.9500

Ball et al. (2006) applied decision trees, trained on the SDSS-DR3
objects with available spectroscopy,
to classify all photometric objects ($> 10^8$) in
SDSS-DR3 in one of the three categories of star, galaxy or nsng
(neither star nor galaxy), the latter including QSOs and `unknown'.
A blind test on the 2dF QSO Redshift Survey (2QZ; Crom et al. 2004), using
the 8739 QSOs matching 2QZ and SDSS-DR3, yielded 95 per cent
completeness and 87 per cent efficiency. The authors do not discuss how
the performance depends on redshift.

Bazell, Miller and SubbaRao (2006) use a semisupervised mixture model
approach to analyze 10000 objects spectrosopically classified in
SDSS-DR4 in the categories of stars, late-type stars, galaxies and
QSOs with $z \le 3$ and unknown, using as input data for the modelling
SDSS colours and the spectroscopic class. Since the aim was to
investigate the existence of possible new object types among the class
of `unknown' as well as subclasses among the remaining classes, 90 per
cent of the sources in the categories of stars, late-type stars,
galaxies and QSOs were also treated as unknown during the modelling.
The best model includes 16 components, two of them of the nonpredefined type,
and one of the latter captures a region of the $u-g$ versus $g-r$
colour-colour diagram ($2 \le u-g \le 5$, $0.5 \le g-r \le 1.5$)
within the location of high-$z$ QSOs in Richards et al. (2002), but
intentionally rejected in the QSO selection by Richards et al. (2004)
because of the high density of stellar contaminants in that region.

Gao, Zhang and Zhao (2008) compare the performance of $k$-dimensional
trees and support vector machines in the separation between stars and
QSOs, using a sample of stars and QSOs spectroscopically classified in
SDSS-DR5 and having a counterpart at the Two-Micron All Sky Survey
(2MASS, Cutri et al. 2003).
Both techniques yield a global efficiency and completeness as large as
97 per cent for $0 \le z \le 2.5$.
However, again the accuracy drops significantly for $z > 2.5$.

Our work deals with the selection of high-$z$ QSOs from SDSS-FIRST matches.
As stated before, our restriction to radio detected sources drastically
reduces the contamination by stars, enabling us to obtain classification
accuracies at these redshifts better than those obtained in more general
studies aimed at the selection of the whole population of QSOs,
regardless of radio detection.
The paper is structured as follows. The sample of FIRST-DR5 matches
is presented in Section 2. In Section 3 we explore the performance
of supervised NNs to separate high-$z$ QSOs from the remaining
spectral classes in the sample of 4250 sources with DR5 spectra,
using multiband optical photometry and radio data. In Sections 4.1
and 4.2, we apply the trained NNs to the sample of 4415 sources
without DR5 spectra, identifying 58 high-$z$ QSO candidates. In
Section 4.3 we check the reliability of this identification via
comparison with spectra from the NASA Extragalactic Database (NED),
SDSS DR6 and follow-up spectroscopy with the WHT. The
discussion and conclusions are presented in Section 5.

\section{SELECTION OF THE SAMPLE}

As an initial sample we selected {\it all} FIRST sources with an
unresolved object in the {\tt PhotoPrimary}\footnote{ Best SDSS
observation of the object, and the object is located within the
imaging survey area which has been finished to date. See
http://cas.sdss.org/astrodr5/en/help/docs/tabledesc.asp} view of the
SDSS DR5 Catalog Archive Server (CAS), within 1.5 arcsec of the
radio position, with dereddened psf magnitude $15 \le r_{\rm AB} <
20.2$ and `clean' photometry (i.e. rejecting objects with magnitude
errors $>$ 0.2 in all five bands $ugriz$, or flagged as `BRIGHT',
`SATURATED', `EDGE', `BLENDED' or `CHILD'). We selected the $r$ band
because QSOs with redshifts $3.6 \le z \le 4.5$ are expected to have
an enhanced emission at this band, due to the Ly$\alpha$ emission
line falling within the covered spectral range. Vigotti et al.
(2003) estimated that more than 99 per cent of FIRST-APM quasars
with $3.8 \le z \le 4.5$, $E \le 18.8$ and $S_{\rm 1.4 ~GHz} > 1.0$
mJy fall within 1.5 arcsec of the POSS I positions, and this
matching radius was adopted for this work. In total 8665 FIRST
sources fulfil the above requirements. Because of the exclusion of
`CHILD' objects (objects which are the product of deblending a
blended object), in all cases there is {\it one} optical object per
radio source. The corrections for Galactic extinction, derived from
Schlegel et al. (1998), were taken from SDSS. 4250 of the sources
(49 per cent of the sample) have DR5 spectra (specifically, they are
included in the {\tt SpecObj}\footnote{This implies that the object
was selected for spectroscopy as an SDSS science object, and that
the spectrum was taken on a main survey plate. See
http://cas.sdss.org/astrodr5/en/help/docs/tabledesc.asp} view of the
DR5 CAS).
%Table 1 lists the FIRST-DR5 overlapping areas and the number
%of matches for both DR5 imaging survey and DR5 spectroscopic survey.
In fact, the magnitude limit $r = 20.2$ was set to ensure an
approximately similar fraction of sources with and without
spectra at DR5.
The distribution of SDSS spectral types and redshifts for the objects
with DR5 spectra, as quoted in {\tt SpecObj}, are given in Table 1.
The redshift distribution of the QSOs is shown in Fig. 1.

The redshifts of the QSOs with $z \ge 3.6$ were checked by visual
inspection of the DR5 spectra.
For two of them, SDSS 130941.36+112540.1 and SDSS 153420.23+413007.5,
we found the redshifts provided by the SDSS pipelines to be likely
incorrect, and subsequently the QSOs were identified
in the SDSS DR5 Quasar Catalog (DR5Q; Schneider et al. 2007)
with revised redshifts $z=1.362$ and $z=1.400$ respectively.
These two QSOs were not considered further and are not included in
Table 1 and Fig. 1.
On the one hand the revised values were published
after we had already trained the NNs with the objects in Table 1
and carried out most of the follow up observations of the selected
candidates. On the other, having been misidentified as high-$z$ QSOs
in DR5 (redshift confidences 0.59 and 0.71 respectively), their exclusion
from the training sample, used for the learning, seems reasonable. Since
our aim is to obtain a high completeness for high-$z$ QSOs, non high-$z$
sources whose spectra can be confused with those of high-$z$ QSOs should
be preferably removed from the training sample, since their inclusion could
hinder the selection of the high-$z$ QSOs whose spectra they resemble.
We note that the redshift for SDSS 153420.23+413007.5 at DR6 has been
updated to the high-confidence manual value $z=1.400$, but for
SDSS 130941.36+112540.1 the redshift obtained with the spectroscopic
pipelines, $z=4.395$ (confidence 0.55), has been maintained at DR6.

\begin{table}
\centering \caption{SDSS classification of the 4248 FIRST-DR5 matches
having SDSS spectra}
\begin{tabular}{lrrr}
\noalign{\smallskip}\hline\noalign{\smallskip}
              Spectral type&Number \\ \hline QSO $z < 3.6$ &3754 \\
              QSO $z \ge 3.6$ & 52 \\
              early-type star & 133 \\
              late-type star & 97 \\
  %            star & 230 \\
              galaxy & 59 \\ unknown & 153 \\ Total & 4248 \\
\noalign{\smallskip}\hline
\end{tabular}
\end{table}

\begin{figure}
\includegraphics[width=82mm]{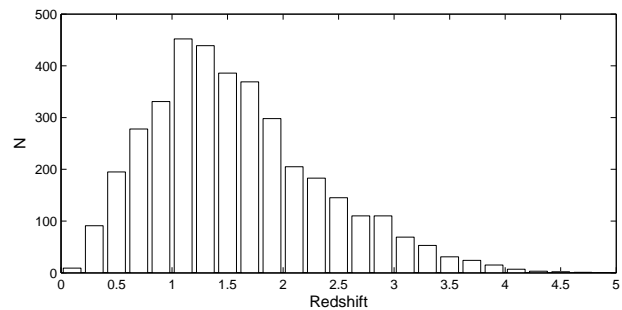}
\caption{Redshift distribution of the 3806 FIRST QSOs with DR5 spectra.}
\end{figure}

\begin{table*}
\centering
\begin{minipage}{125mm}
\caption{
FIRST-DR5 sources with SDSS spectra, and classified by SDSS as
$z \ge 3.6$ QSOs
}
\begin{tabular}{cccrcccc}
\hline
    RA   &Dec    &$r_{\rm AB}$    &$S_{\rm 1.4 ~GHz}$&Redshift &NN output &Previous &Notes\\
    J2000&J2000  &   &        mJy~~     &         &$y_{\rm med}$       &samples  & \\
         (1)&(2)&(3)&(4)~~~&(5)&(6)&(7)&(8)\\\hline
  01 53 39.61 &  -00 11  05.0&  18.82 &   4.75 & 4.194 &  1.00   &       &     \\
  03 00 25.22 &   00 32  24.2&  19.66 &   7.56 & 4.201 &  1.00   &       &     \\
  07 51 13.05 &   31 20  37.9&  19.75 &   5.60 & 3.761 &  0.11   &       & abs \\
  07 51 22.35 &   45 23  34.2&  20.20 &   1.13 & 3.608 &  0.89   &       &     \\
  08 10 09.95 &   38 47  57.0&  19.62 &  27.16 & 3.946 &  0.29   &       &     \\
  08 38 08.46 &   53 48  09.8&  19.94 &   8.42 & 3.610 &  1.00   &       &     \\
  08 39 46.22 &   51 12  02.8&  19.33 &  41.64 & 4.390 &  1.00   & 1,2   &     \\
  08 40 44.18 &   34 11  01.6&  19.79 &  13.59 & 3.889 &  1.00   &       &     \\
  08 52 57.12 &   24 31  03.1&  19.47 & 159.58 & 3.617 &  0.33   &       &     \\
  09 18 24.38 &   06 36  53.3&  19.77 &  26.50 & 4.192 &  0.84   & 2     & abs \\
  09 37 14.49 &   08 28  58.5&  18.59 &   3.17 & 3.700 &  0.53   &       &     \\
  09 40 03.02 &   51 16  02.7&  19.99 &  13.91 & 3.601 &  0.03   & 3     &     \\
  10 00 12.26 &   10 21  51.8&  19.54 &  21.93 & 3.638 &  0.86   &       &     \\
  10 17 47.75 &   34 27  37.8&  20.00 &   2.63 & 3.691 &  0.91   &       &     \\
  10 30 55.95 &   43 20  37.7&  19.84 &  37.82 & 3.700 &  0.75   &       &     \\
  10 34 46.54 &   11 02  14.5&  18.80 &   1.09 & 4.266 &  0.70   &       &     \\
  10 51 21.36 &   61 20  38.0&  18.90 &   6.64 & 3.689 &  0.93   &       &     \\
  10 57 56.28 &   45 55  53.0&  17.44 &   1.38 & 4.137 &  1.00   & 1,2,3 &     \\
  11 10 55.21 &   43 05  10.0&  18.59 &   1.21 & 3.821 &  0.81   & 3     &     \\
  11 17 01.89 &   13 11  15.4&  18.28 &  28.99 & 3.624 &  0.14   &       &     \\
  11 17 36.33 &   44 56  55.6&  20.03 &  25.08 & 3.853 &  1.00   &       &     \\
  11 25 30.48 &   57 57  22.7&  19.41 &   2.99 & 3.685 &  0.70   &       & abs \\
  11 27 49.45 &   05 11  40.6&  19.13 &   2.71 & 3.711 &  0.12   &       & abs \\
  11 29 38.73 &   13 12  32.2&  18.77 &   1.33 & 3.607 &  0.21   &       &     \\
  11 33 30.91 &   38 06  38.1&  19.71 &   0.87 & 3.631 &  0.80   &       &     \\
  11 50 45.61 &   42 40  01.1&  19.87 &   1.51 & 3.894 &  0.54   &       &     \\
  12 04 47.15 &   33 09  38.7&  19.24 &   0.92 & 3.616 &  0.56   &       & BAL \\
  12 31 42.17 &   38 16  58.9&  20.18 &  24.04 & 4.138 &  0.99   &       &     \\
  12 40 54.91 &   54 36  52.2&  19.74 &  15.09 & 3.938 &  1.00   & 3     &     \\
  12 42 09.81 &   37 20  05.6&  19.34 & 662.38 & 3.819 &  0.97   &       &     \\
  12 46 58.83 &   12 08  54.7&  20.00 &   1.44 & 3.805 &  1.00   &       & BAL \\
  12 49 43.67 &   15 27  07.0&  19.31 &   2.01 & 3.995 &  0.89   &       &     \\
  13 00 02.16 &   01 18  23.0&  19.78 &   2.52 & 4.614 &  0.90   &       &     \\
  13 03 48.94 &   00 20  10.4&  18.90 &   0.99 & 3.647 &  0.99   &       & BAL \\
  13 07 38.83 &   15 07  52.1&  19.70 &   3.89 & 4.082 &  0.94   &       &     \\
  13 15 36.57 &   48 56  29.1&  19.77 &  10.86 & 3.618 &  0.96   &       &     \\
  13 25 12.49 &   11 23  29.7&  19.33 &  71.05 & 4.409 &  1.00   & 2     &     \\
  13 54 06.89 &  -02 06  03.2&  19.18 & 719.48 & 3.715 &  0.67   &       &     \\
  13 55 54.55 &   45 04  21.0&  19.34 &   2.07 & 4.095 &  1.00   &       &     \\
  14 08 50.91 &   02 05  22.7&  19.07 &   1.18 & 4.008 &  0.06   &       & abs \\
  14 12 09.96 &   06 24  06.8&  20.17 &  43.47 & 4.467$^a$ &  0.90   &       & abs \\
  14 22 09.70 &   46 59  32.5&  19.70 &  11.03 & 3.798 &  1.00   &       &     \\
  14 23 26.48 &   39 12  26.2&  20.15 &   6.52 & 3.921 &  1.00   &       &     \\
  14 35 33.77 &   54 35  59.3&  20.05 &  95.78 & 3.810 &  0.90   &       &     \\
  14 45 42.75 &   49 02  48.9&  17.32 &   3.18 & 3.876 &  0.99   &       &     \\
  14 46 43.36 &   60 27  14.3&  19.79 &   1.81 & 3.777 &  1.00   &       &     \\
  15 03 28.88 &   04 19  49.0&  17.96 & 124.97 & 3.664 &  0.84   &       &     \\
  15 06 43.81 &   53 31  34.3&  18.94 &  14.63 & 3.790 &  0.32   & 3     & abs \\
  16 17 16.49 &   25 02  08.1&  19.99 &   2.35 & 3.943 &  0.98   &       & BAL \\
  16 19 33.65 &   30 21  15.1&  19.52 &   4.19 & 3.794 &  0.92   & 3     &     \\
  16 39 50.52 &   43 40  03.7&  17.96 &  25.23 & 3.990 &  0.20   & 1,2,3 & abs \\
  16 43 26.24 &   41 03  43.5&  20.10 &  65.01 & 3.873 &  1.00   &       &     \\
\hline
\end{tabular}

The columns give: (1,2) SDSS coordinates; (3) SDSS dereddened psf $r_{\rm AB}$
magnitude; (4) FIRST total radio flux density; (5) SDSS redshift
($a$ = revised redshift from DR5Q, see Section 2) ; (6)
NN output (see Section 4.2); (7) labels 1, 2 and 3 indicate QSOs
included in the samples of Benn et al. (2002), Holt et
al. (2004) and Carballo et al. (2006) respectively; (8)
BAL = broad absorption line QSO; abs = the Ly$\alpha$ line
appears to be self-absorbed.

 \end{minipage}
 \end{table*}

The spectral types and redshifts of the remaining 4196 sources in Table 1,
most of them QSOs at $z < 3.6$, were taken directly from DR5, since
we did not expect among them any $z \ge 3.6$ QSO
with an identification as reliable as that found for the 52 high-$z$ QSOs.
A visual examination of the DR5 spectra of 225 of these sources
(the first 125 $z < 3.6$ QSOs,
25 early-type stars, 25 late-type stars, 25 galaxies, and 25 `unknown',
in order of increasing right ascension) yielded no identification as a
likely $z \ge 3.6$ QSO. Moreover, since our approach for the automated
selection of high-$z$ QSOs
groups the remaining spectral types as a single class (see Section 3),
revisions from DR5 such as shifts in redshift below this limit or changes
between the non high-$z$ QSO categories would not affect the results.

The fraction of QSOs with $z \ge 3.6$ is 1.37\% (52/3806) and they are
listed in Table 2. All these QSOs are included in DR5Q and only one of them,
SDSS 141209.96+062406.8, with a deep and wide absorption feature bluewards
of the Ly$\alpha$ emission line and starting at the Ly$\alpha$ line,
has a revised redshift at DR5Q, $z=4.467$ versus $z=4.421$ at DR5.
DR5Q provides interesting complementary information for these QSOs and for
the remaining ones in Table 1, including $i$ band absolute magnitudes, $g-i$
`differential colour' with respect to the typical value for the QSO redshift,
and matches to $ROSAT$ All-Sky Survey
(RASS; Voges et al. 1999, 2000) and 2MASS when available.

%Fig. 1 shows the redshift distribution of the QSOs in Table 1.
%The fraction of QSOs with $z \ge 3.6$ is 1.37\% (52/3806).
%These 52 $z \ge 3.6$ QSOs are listed in Table 2.

The FIRST-SDSS sample was obtained using a simple one-to-one match between
radio and optical sources (within a 1.5 arcsec radius), therefore missing
the class of double-lobe QSOs without detected radio cores.
Using the statistics found by
de Vries et al. (2006, their table 2) for a sample of 5515 FIRST-SDSS QSOs
with radio morphological information within 450 arcsec, the fraction of
FIRST-SDSS double-lobe QSOs with undetected cores with respect to the
total FIRST-SDSS QSO sample is 3.7 per cent. Since the starting samples
of SDSS QSOs in de Vries et al. (2006) and in this work obbey similar
SDSS selection criteria, the last value is a good estimate of the
incompleteness of the QSO samples in our work due to the exclusion of
lobe dominated QSOs.

\section{SEPARABILITY OF HIGH-REDSHIFT QSOS WITH A NEURAL NETWORK}

\begin{table}
\centering \caption{Combinations (A - H) of optical and radio data
used as input variables to the neural network}
\begin{tabular}{lrrrrrrrr}
\noalign{\smallskip}\hline\noalign{\smallskip}
                   &A&B&C&D&E&F&G&H\\
\hline
 $r$ magnitude                    &$\times$&$\times$&$\times$&$\times$&$\times$&$\times$&$\times$&$\times$\\
 $u-g$                                  &$\times$&$\times$&$\times$&$\times$&$\times$&$\times$&$\times$&$\times$\\
 $g-r$                                   &$\times$&$\times$&$\times$&$\times$&$\times$&$\times$&$\times$&$\times$\\
 $r-i$                                   &$\times$&$\times$&$\times$&$\times$&$\times$&$\times$&$\times$&$\times$\\
 $i-z$                                     &$\times$&$\times$&$\times$&$\times$&$\times$&$\times$&$\times$&$\times$\\
 rad-opt separation                            & &$\times$& & &$\times$&$\times$& &$\times$\\
log$_{\rm 10}$$[S_{\rm total}$(1.4 GHz)]       & & &$\times$& &$\times$& &$\times$&$\times$\\
 log$_{\rm 10}$$(S_{\rm total}/S_{\rm peak})$  & & & &$\times$& &$\times$&$\times$&$\times$\\
\noalign{\smallskip}\hline
\end{tabular}
\end{table}

Only 52 of the 4248 sources with DR5 spectra are classified as $z
\ge 3.6$ QSOs (Table 1). Our goal is to train a classifier to
recognize high-$z$ QSOs among the 4415 objects {\it without} SDSS
spectra, i.e. the `unlabelled' sources, after learning the class
properties from the 4248 objects {\it with} spectra, i.e. the
`labelled' sources. The adopted procedure was to consider a
two-class problem, with high-$z$ QSOs as one class and the remaining
types as the other. Since the training uses objects whose
class is known, the learning is said to
be `supervised'.

Previous selections of high-$z$ RL QSOs as FIRST sources with
optical counterparts on POSS-APM revealed an abrupt change in $O-E$
colour with redshift at $\sim 3.6$, the latter allowing
efficient separation of high-$z$ QSOs from the QSO population as a
whole (Benn et al. 2002, Vigotti et al. 2003, Carballo et al. 2006).
We therefore took this redshift as an initial threshold for high-$z$
QSOs, although we explored redshifts below $z=3.6$ to find the
optimal value for the optical bands used in this work.

The learning algorithm applied was a feed-forward NN (Bishop 1995)
with a layer for the input parameters, i.e. the data, and an output
layer with a single variable $y$, set during training to 1 for
high-$z$ QSOs and 0 for the remaining types. Output $y$ for object
$i$ is given by:

$$y^i = f(a^i ) =  {1 \over 1 + e^{-a^i}}~, $$

$${\rm with ~~~~} a^i = w_0 + \sum_{j=1}^d w_j \cdot x_j^i~,$$

\noindent where $(x_1, x_2,  \ldots, x_d)$ are the inputs,
$a$ is a linear function of the inputs, and $f$ is the non-linear
function known as a
logistic sigmoid, with outputs in the range $(0,1)$. This NN model is known
as logistic linear discriminant.
$w_0$ and $(w_1, w_2, \ldots, w_d)$, called bias and weights
respectively, are the parameters fitted during training. The adopted error
function was the mean of the squared errors of the outputs,

 $$mse = {1 \over m} \sum_{i=1}^{m} (y^i - t^i)^2,$$

\noindent where $m$ is the number of objects for the training and
$t$ is the desired value of output $y$ or target value. The optimal
parameters for the net, i.e.  those minimising the error, were
obtained using the Levenberg-Marquardt algorithm, available in the
MATLAB Neural Network Toolbox (http://www.matchworkds.com/). The
Levenberg-Marquardt algorithm appears to be the fastest method for
training moderate-sized NNs (Hagan \& Menhaj 1994),
and its efficient implementation in Matlab further improves its performance.

As input data we tried various combinations (A - H) of variables
shown in Table 3. A pre-processing was performed normalising each
variable to the range $[-1,1]$. For this step the total sample was
used (unlabelled sources included) since the inputs for the new
objects presented to the net need to have the same normalization as
the data used in the learning process. The output values, $0 \le y
\le 1$, give the degree of similarity with the class of high-$z$
QSOs. Objects with $y$ exceeding a given threshold $y_{\rm c}$ would
be classified as high-$z$ QSO candidates.

The quality of an NN for classification is evaluated in terms of its
efficiency, {\it eff}, and completeness, {\it comp}.
Efficiency is
the fraction of sources with $y \ge y_{\rm c}$ that actually are
high-$z$ QSOs. Completeness is the fraction of actual high-$z$ QSOs
with $y \ge y_{\rm c}$.
A good separation between two classes will show
efficiency increasing and completeness decreasing as
$y_c$ increases.
Since our purpose is to build a sample appropriate
for statistical analysis, priority is given to completeness,
accepting low $y_c$ values at the cost of lower efficiency.

The classifier has to be empirically tested using a set of
objects not used for the learning. Because of the small
number of high-$z$ QSOs, the training and test samples were
separated adopting the partition method known as `leave-one-out',
repeatedly dividing the data set of $m$ instances into a training
set of size $m-1$ and a test set of size 1, in all possible ways.
This procedure yielded $m$ classifiers, one per test object. Since
the $m$ objects (4248) are used for testing, a good estimate of
the performance can be obtained.

\begin{figure}
\includegraphics[width=82mm]{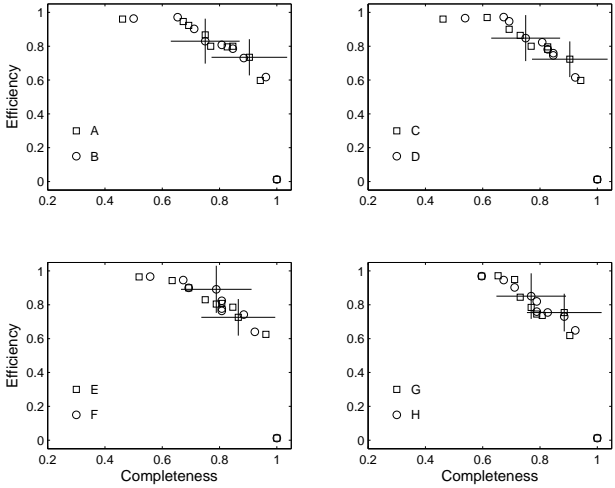}
\caption{Efficiency versus completeness of the NN search for
high-$z$ QSOs, for $y_c$ values $0,0.1,0.2,0.3,0.4,0.5,0.6,0.7,0.8,0.9$
(right to left) and the eight sets of input variables A--H.
The adopted redshift cut for high-$z$ QSOs was $z=3.6$.
A sample poisson error bar is plotted for each set.}
\end{figure}

 \begin{figure} \includegraphics[width=82mm]{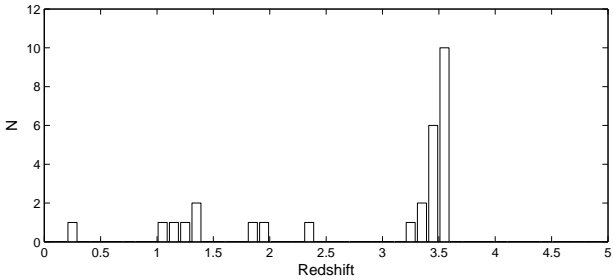}
 \caption{
 %Top panel: Similar to Fig. 2, but for set of inputs B (the
 %best-performing set of inputs). Lower panel:
 Redshift distribution for the 28 $z < 3.6$ QSOs with $y \ge 0.1$
(i.e. the contaminants), using the best-performing set of inputs, B.}
 \end{figure}

\begin{figure}
\includegraphics[width=82mm] {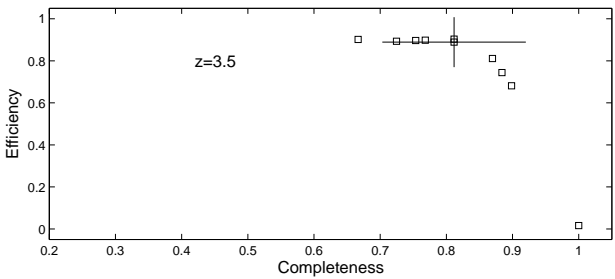} \caption{Similar to Fig. 2,
but using the set of inputs B and with $z_{\rm cut}=3.5$. The number
of QSOs with higher redshift than this is 69.}
\end{figure}

The efficiency and completeness as a function of $y_{\rm c}$ for each
of the eight sets of input variables is shown in Fig. 2. For all sets
of features the expected trend between {\it eff} and {\it comp}
is present, and classifiers are found
yielding {\it comp} $\ge$ 90 per cent and {\it eff} $\ge$ $60$ per
cent for particular $y_c$ values. Among these sets of inputs, B and
E gave the maximum completeness, $\sim$ 96 per cent, both yielding
$\sim$ 62 per cent efficiency.  B was selected as the best
classifier since it uses a smaller number of input variables.
%A plot similar to Fig. 2 for this classifier is shown in Fig. 4.
In total, 81 sources have $y \ge 0.1$ for this set, of which 50 are
QSOs with $z \ge 3.6$, yielding completeness 50/52 = $96 \pm 14$ per
cent and efficiency 50/81 = $62 \pm 9$ per cent for $y \ge 0.1$.
 The 31 contaminants include one star, two galaxies and 28 $z < 3.6$ QSOs.
The redshift distribution of the latter is shown in Fig. 3.
19 of the QSOs, i.e. a fraction $19/31 = 61 \pm 14$ per
cent of the contaminants, have redshifts $3.2 \le z < 3.6$, very
close to the selected threshold.

Using a redshift cut $z_{cut}=3.5$ and the set of inputs B we found
$comp = 90 \pm 11$ per cent and {\it eff} = $68 \pm 9$ per cent for
$y_c=0.1$. (Fig. 4).
Since completeness was prioritised we kept $z_{\rm
cut}=3.6$, yielding completeness $\sim 96$ per cent.

The inputs used for the learning were basically the SDSS colours,
which are known to be correlated, especially for QSOs (Weinstein et
al. 2004).  This means that we could have applied some preprocessing
algorithms to reduce the input space dimension, and therefore improve
its sampling. In fact, the approach of logistic linear regression does
not need to assume that the variables are independent, and the presence
of covariance in the input data does not affect the
quantification of the parameters of the optimal hyperplane separating
high-$z$ QSOs from the remaining classes except for causing
a sampling of the input space lower than necessary.
The good performance found for the test set
of unseen data gives us confidence that although probably redundant,
the selected set of input features is appropriate.

Several works described in the Introduction select QSOs from SDSS
using colour and/or photometric information covering the five bands.
Suchkov et al. (2005) apply
a $k$-dimensional decision tree classifier and use five
colours ($u-g$, $g-r$, $r-i$, $i-z$ and $g-i$) as input data.
Gao et al. (2008) apply decision trees and
support vector machines to select QSOs from a combined SDSS-2MASS
sample, using several sets of input data including in all cases
SDSS photometry at the five bands (magnitudes in all bands
or four colours and a magnitude). They
found the best results for the input set with four SDSS coulours
and the $r$-band magnitude, which is the set of optical data
selected in our work.
Ball et al. (2006, 2007) use decision trees to select QSOs from SDSS
and a $k$-nearest neighbour instance-based approach to quantify QSO
photometric redshifts (Ball et al. 2008), in both cases using as
input data four colours in the four magnitude types
(PSF, fiber, Petrosian and model).  Ball et al. (2007) applied genetic
algorithms to investigate subsets of the original 16 inputs in a
systematic way and found that no subset resulted in a significant
improvement and some subsets were even worse, therefore electing
to keep the full SDSS information available.

Since we adopted the leave-one-out approach and use all the sources as test,
training and testing objects form the same sample, i.e. the sample
of sources with DR5 spectra.
This means that the quoted figures of $96 \pm 14$ completeness
and $62 \pm 9$ per cent efficiency refer to the selection of high-$z$ QSOs
among the labelled sources, i.e. within the interpolation regime.

\begin{figure*}
\includegraphics[width=179mm]{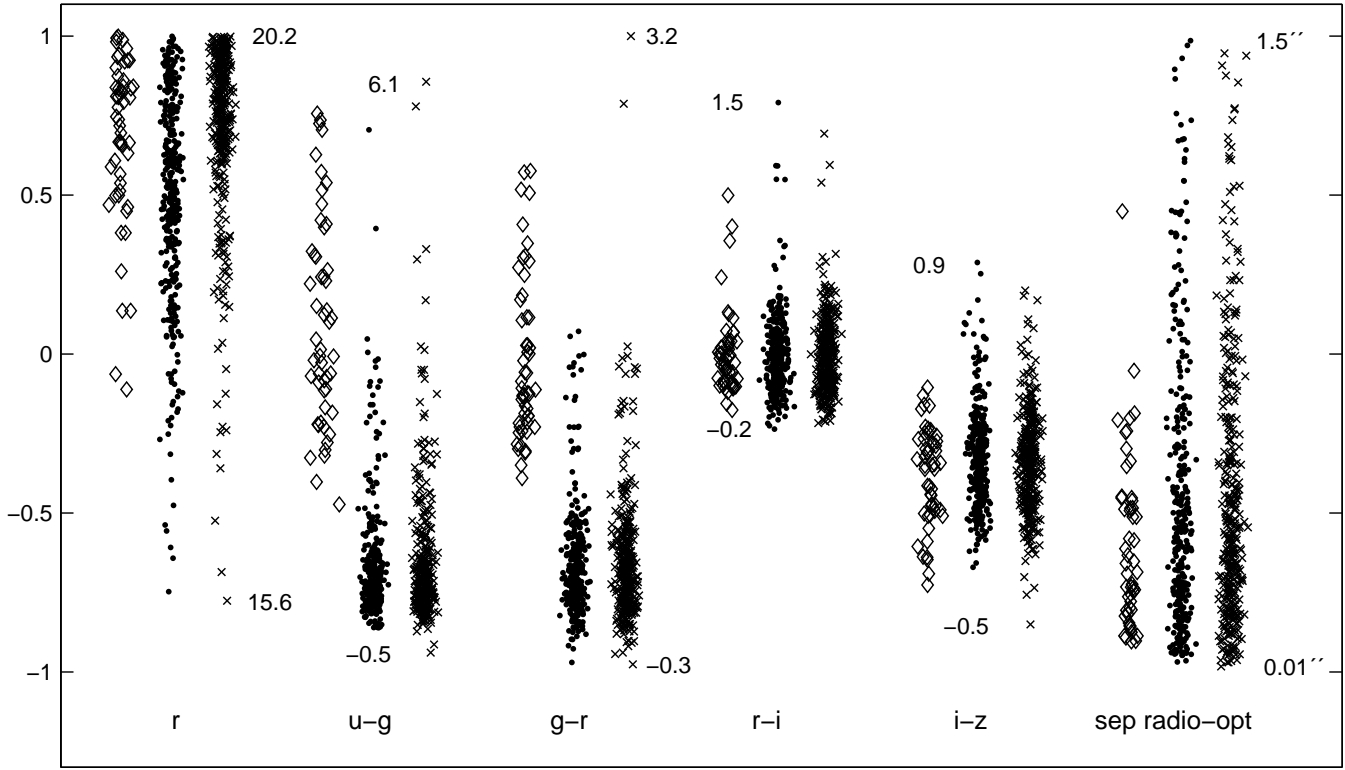} \caption{ Normalized
input parameters $r$, $u-g$, $g-r$, $r-i$, $i-z$
and radio-optical separation for labelled high-$z$ QSOs ($\diamond$),
labelled non high-$z$ QSOs ($\bullet$, 8 per cent shown),
and unlabelled sources ($\times$, 8 per cent shown).
The numbers indicate the minimum and maximum value of each variable
(in magnitudes or arcsec) for the sources in the plot.}
\end{figure*}

The labelled sample contains, among the 52 high-$z$ QSOs,
12 sources with broad absorption lines (BALs) or self-absorption at
Ly$\alpha$ (see  Table 2). The classifier
recovers 50/52 of the high-$z$ QSOs and 11/12 of the BALs and self-absorbed
at Ly$\alpha$, proving to be effective in the selection of QSOs
with this type of absorption features.
On the other hand, we confirmed that all 52
high-$z$ QSOs in the training sample have at least an emission line with
FWHM $\ge$ 1000 km sec$^{-1}$, i.e. none of them belongs to the class
of ``narrow-lined'' (type II) QSOs. Therefore the classifier trained in this
work targets the selection of high-$z$ QSOs with at least a broad emission
line, including those cases presenting absorption features in the form of BALs
or Ly$\alpha$ absorption.

Figure 5 shows the distribution of the six variables used as
input data $r$, $u-g$, $g-r$, $r-i$, $i-z$ and radio-optical offset,
separating (from left to right) labelled high-$z$ QSOs,
labelled non high-$z$ QSOs and unlabelled sources.
For non high-$z$ QSOs (4196) and unlabelled sources (4415) we used for
the plot representative subsets containing
eight per cent of the sources, to improve visualization.
For non high-$z$ QSOs this proportion was
applied separately for QSOs, galaxies, early-type stars, late-type stars and
`unknown', to keep the same fractions as in the original sample.
 %, yielding 300, 8, 11, 5 and 12 sources respectively.
Most of the sources in this sample are QSOs with $z<3.6$ (3906/4248
= 90 per cent). The scale is linear, and although normalized
variables were used, their ranges in physical units can be inferred
from the marked numbers in the figure, showing the minimum and
maximum value of each variable for the represented sources. Table 4
presents the mean, standard deviation and median for each variable,
for the complete three samples, as well as for the whole sample of
labelled sources. In Fig. 5 and Table 4 all sources were included
regardless of their photometric errors. The errors at $g$, $r$, $i$
and $z$ are less than 0.2 mag (except for one unlabelled source with
$\Delta r=0.24$ and eight with $\Delta z = 0.2 - 0.3$). For the $u$
band the errors exceed 0.2 mag for all high-$z$ QSOs, 246 labelled
non high-$z$ QSOs (out of 4196) and 313 unlabelled sources (out of
4415), with median values for these sources of $\Delta u=0.8, 0.5,
0.4$ respectively.

\begin{table}
\centering \caption{Statistics (mean, standard deviation and median) of the selected input variables for various SDSS-DR5 subsamples}
\begin{tabular}{lllll}
\noalign{\smallskip}\hline\noalign{\smallskip}
                   & high-$z$ &non high-$z$    &lab.      &unlab.           \\
                   & QSOs     &QSOs            &          &                 \\
                   & (52)     &(4196)          &(4248)    &(4415)           \\
\hline
$r$   &19.36$\pm$0.69&18.90$\pm$0.86&18.90$\pm$0.86    & 19.33$\pm$0.81  \\
         &19.53         &19.00         &19.01             & 19.56           \\
$u$$-$$g$    & 3.34$\pm$1.20& 0.54$\pm$0.75& 0.57$\pm$0.82    &  0.53$\pm$0.70  \\
         & 3.07         & 0.31         & 0.32             &  0.32           \\
$g$$-$$r$    & 1.40$\pm$0.45& 0.27$\pm$0.31& 0.29$\pm$0.33    &  0.27$\pm$0.30  \\
         & 1.24         & 0.21         & 0.21             &  0.21           \\
$r$$-$$i$    & 0.17$\pm$0.22& 0.16$\pm$0.21& 0.16$\pm$0.22    &  0.15$\pm$0.20  \\
         & 0.13         & 0.14         & 0.14             &  0.13           \\
$i$$-$$z$    & 0.06$\pm$0.18& 0.11$\pm$0.17& 0.11$\pm$0.17    &  0.12$\pm$0.18  \\
         & 0.10         & 0.09         & 0.09             &  0.11           \\
sep('')  & 0.30$\pm$0.20& 0.40$\pm$0.33& 0.40$\pm$0.33    &  0.39$\pm$0.32  \\
     & 0.25         & 0.29         & 0.29           &  0.28           \\
\noalign{\smallskip}\hline
\end{tabular}
\end{table}

Regarding the comparison among labelled sources, the variables that
taken individually better discriminate between the classes of high-$z$
QSOs and non high-$z$ QSOs are the $u-g$ and $g-r$ colours, as expected
from the fact that $z<3.6$ QSOs dominate the non high-$z$ QSO sample and
the well established colour-redshift relation for QSOs (e.g.
Schneider et al. 2007). Also noticeable
is the concentration of high-$z$ QSOs at the faintest $r$ magnitudes
(fainter magnitude and lower dispersion) and at the lowest radio-optical
offsets (lower separation and again lower dispersion), compared to the
remaining labelled sources.

\section{APPLICATION OF THE NETWORK TO THE UNLABELLED SAMPLE}

\subsection{The labelled and unlabelled samples}

The labelled sample consists of the FIRST-SDSS sources included in
the DR5 spectroscopic catalogue. The content of the labelled sample
is determined by the way the photometric objects were selected as
spectroscopic targets by SDSS. The selection criteria were mainly
aimed at obtaining samples of galaxies, QSOs and brown dwarfs, with
different combinations of selected parameters (e.g. magnitude and
colour ranges, extension, proximity to catalogued sources at other
wavelengths) being used for each object type. The labelled sample
cannot therefore be considered as statistically representative of
the SDSS imaging database. Classes not considered in the
spectroscopic selection criteria may be absent or poorly represented
in the spectroscopic catalogue. The unlabelled sample is therefore a
mixture of the sources in the DR5 spectroscopic area not selected as
spectroscopic targets (2059 objects within $\sim$ 5553 deg$^2$,
compared to 4248 labelled in the same region), and the sources in
the DR5 photometric area but outside the spectroscopic area (2356
objects within $\sim$ 1838 deg$^2$).

A general concern about classification is the application of an
algorithm trained on a sample of data to a different set of data,
likely covering other regions of input space. The extension of
the classifier beyond the original sample used for the training
is framed in the context of extrapolation versus interpolation.
In our
case the training set is the DR5 spectroscopic survey or
labelled sample, and our aim is to apply the classifier to the
sources in the DR5 photometric sample without DR5 spectra.
We expect a reasonable overlap between labelled sources and the
sources located outside the spectroscopic area, since a large
fraction of the latter would have been SDSS spectroscopic targets if
included in the spectroscopic area [a fraction 4248/(4248+2059) = 67
per cent using the statistics in DR5 spectroscopic area]. A poorer
overlap is expected between labelled sources and the unlabelled
sources in DR5 spectroscopic area.

Figure 5 and Table 4 allow to compare the distribution of
each input variable for the classes of labelled and unlabelled
sources. Although Fig. 5 does not include labelled sources as a
whole, the distributions for this set are approximately similar
to those for labelled non high-$z$ QSOs, since high-$z$
QSOs make only 1.2 per cent (52/4248) of the labelled sources.
The main effect of including high-$z$ QSOs would be to increase
the ranges of the $u-g$ and $g-r$ colours.
The statistics of mean, standard deviation and median for the whole
labelled and unlabelled groups are given in Table 4 (columns 3 and 4).
Figure 5 and Table 4 show a good agreement between the distributions
of each of the input variables for unlabelled and labelled sources, except
for the $r$ band magnitudes, which are fainter
for unlabelled sources, with mean and median of 19.33 and 19.56 versus
18.90 and 19.01.
These comparisons between unlabelled and labelled sources use each
input variable individually.
More complex separations between labelled and unlabelled
sources would be expected in the six-dimensional input space.

We applied the trained classifier to the unlabelled sample in order
to select new high-$z$ QSO candidates and to obtain an estimate of
the completeness of SDSS spectroscopy for radio QSOs with
$3.6 \le z \le 4.6$.
The sample of new high-$z$ QSO candidates is presented
in Sect. 4.2. In Sect. 4.3 we discuss the performance obtained in
the unlabelled sample, on the basis of available spectroscopic
identifications from observations in this work, taken from the
literature or from SDSS DR6.

\begin{figure}
\includegraphics[width=84mm]{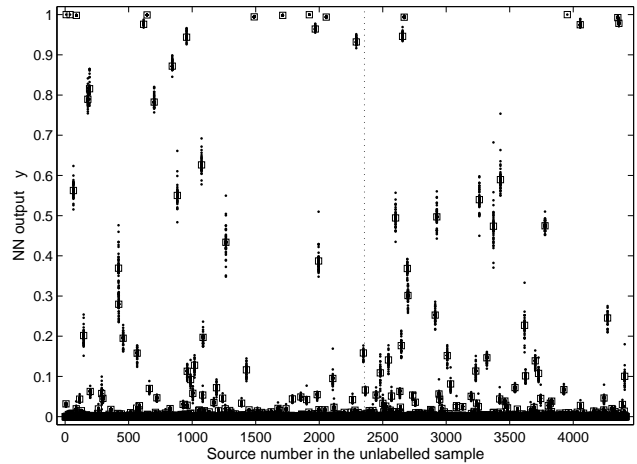}
\caption{Distribution of NN output $y$ for the unlabelled FIRST-DR5
sources and the separation high-$z$ QSO versus remaining classes,
assuming the trained NNs (see Section 4.2). The boxes mark the
median values. The sources to the left of the vertical line (2356) are
located outside the DR5 spectroscopic area, and those to right (2059)
inside this area.}
\end{figure}

\begin{table*}
\centering
\begin{minipage}{160mm}
\caption{$z \ge 3.6$ QSO candidates identified by the NN in the FIRST-SDSS DR5 unlabelled sample}
\begin{tabular}{cccrcllc}
\hline
RA   &Dec  &$r_{\rm AB}$& $S_{\rm 1.4 ~GHz}$&NN output &\hspace{0.6cm}Redshift&\hspace{0.8cm}Notes &In DR6       \\
J2000&J2000&            &         mJy~~     &$y_{\rm med}$&NED\hspace{0.2cm}WHT\hspace{0.2cm}DR6&  &spec. area?\\
         (1)&(2)&(3)&(4)~~~&(5)&\hspace{0.9cm}(6)&\hspace{0.9cm}(7)&(8)\\\hline
 07  25  18.26  &  37  05  18.3 &19.61 &   26.56 &   1.00 &\hspace{0.2cm}4.33&                    &     \\
 07  47  11.15  &  27  39  03.3 &18.35 &    1.55 &   0.11 &\hspace{0.2cm}4.17&                    &yes  \\
 07  47  38.49  &  13  37  47.3 &19.37 &    6.62 &   1.00 &\hspace{1.0cm}4.04&BAL                 &  \\
 08  07  10.74  &  13  17  39.4 &20.01 &   48.88 &   0.56 &\hspace{1.0cm}3.70\hspace{0.3cm}3.726& &yes     \\
 08  15  55.01  &  46  53  21.4 &19.88 &    3.73 &   0.14 &    &                                  &yes  \\
 08  23  23.32  &  15  52  06.8 &19.31 &   79.33 &   1.00 &\hspace{1.0cm}3.79\hspace{0.3cm}3.781& &yes  \\
 08  33  16.91  &  29  22  28.0 &20.13 &   12.73 &   0.49 &    &                                  &yes  \\
 08  43  23.69  &  16  56  56.1 &19.66 &    2.36 &   0.20 &    &                   DR6 unknown    &yes  \\
 08  48  18.88  &  39  38  06.0 &20.15 &    0.72 &   0.18 &    &                                  &yes  \\
 08  52  58.87  &  22  50  50.5 &20.16 &   45.33 &   0.95 &\hspace{1.0cm}3.55&                    &yes  \\
 08  55  01.82  &  18  24  37.8 &19.96 &    9.43 &   0.79 &\hspace{1.0cm}3.96\hspace{0.3cm}3.966& &yes  \\
 08  59  44.06  &  21  25  11.1 &18.72 &   23.96 &   0.82 &\hspace{1.0cm}3.70\hspace{0.3cm}3.699& &yes  \\
 09  02  54.17  &  41  35  06.5 &20.12 &    0.93 &   0.99 &\hspace{1.0cm}3.69&                    &yes  \\
 09  09  53.84  &  47  49  43.0 &19.89 &  383.67 &   0.37 &    &                                  &yes  \\
 09  14  36.22  &  50  38  48.5 &20.16 &   51.00 &   0.30 &    &                                  &yes  \\
 10  09  33.22  &  25  59  01.2 &20.13 &    3.42 &   0.37 &    &                   DR6 unknown    &yes  \\
 10  10  20.85  &  28  51  50.1 &20.11 &    2.62 &   0.28 &    &DR6 $z=0.58$ galaxy&yes  \\
 10  19  39.00  &  19  03  12.0 &20.11 &    0.74 &   0.20 &    &DR6 $z=0.45$ galaxy&yes  \\
 10  29  40.93  &  10  04  10.9 &19.47 &    3.22 &   0.25 &    &                   &yes  \\
 10  34  20.43  &  41  49  37.5 &20.12 &    1.98 &   0.50 &    &                   &yes  \\
 10  52  25.06  &  25  15  41.3 &20.10 &    5.26 &   0.16 &\hspace{1.9cm}3.404&    &yes  \\
 10  58  07.47  &  03  30  59.6 &19.92 &    4.62 &   0.15 &    &                   &yes  \\
 11  05  43.86  &  25  53  43.1 &20.09 &    2.76 &   0.98 &\hspace{1.9cm}3.779&    &yes  \\
 11  09  46.44  &  19  02  57.6 &20.04 &    6.95 &   1.00 &    &                   &     \\
 11  23  39.59  &  29  17  10.7 &19.47 &    3.14 &   0.78 &\hspace{1.9cm}3.771&    &yes  \\
 11  46  41.02  &  12  52  02.9 &20.19 &    3.01 &   0.11 &    &                   &yes  \\
 11  51  07.42  &  50  15  58.5 &20.08 &    1.22 &   0.54 &    &                   &yes  \\
 11  54  49.36  &  18  02  04.4 &19.63 &   39.06 &   0.87 &\hspace{1.9cm}3.688&    &yes  \\
 12  04  07.83  &  48  45  48.2 &19.97 &    2.64 &   0.15 &    &                   &yes  \\
 12  05  31.73  &  29  01  49.2 &20.17 &    1.46 &   0.55 &    &                   &yes  \\
 12  13  29.42  & -03  27  25.7 &19.64 &   25.53 &   0.47 &    &                   &yes  \\
 12  20  27.96  &  26  19  03.5 &18.12 &   35.04 &   0.94 &\hspace{0.2cm}3.694\hspace{1.0cm}3.697&&yes \\
 12  21  35.64  &  28  06  13.8 &19.77 &   28.76 &   0.11 &\hspace{0.2cm}3.305\hspace{1.0cm}3.288&&yes  \\
 12  28  19.96  &  47  40  30.4 &19.32 &    2.22 &   0.59 &    &                   &yes  \\
 12  31  28.22  &  18  47  14.3 &19.41 &   11.17 &   0.13 &    &                   &     \\
 12  43  23.16  &  23  58  42.2 &19.87 &   63.44 &   0.63 &    &                   &     \\
 12  44  43.06  &  06  09  34.6 &19.78 &    1.36 &   0.20 &    &                   &yes  \\
 13  12  42.86  &  08  41  05.0 &18.53 &    3.93 &   0.43 &\hspace{1.0cm}3.73\hspace{0.3cm}3.743&&yes\\
 13  20  53.63  &  10  37  51.5 &19.46 &    8.43 &   0.23 &\hspace{1.0cm}3.42\hspace{0.3cm}3.431&&yes\\
 13  22  27.58  &  53  52  09.2 &19.68 &    2.51 &   0.10 &\hspace{1.0cm}1.23 &   &yes \\
 13  37  59.43  &  36  34  20.6 &20.17 &    2.88 &   0.12 &\hspace{1.0cm}1.07 &   &yes \\
 13  42  01.42  &  05  01  56.0 &20.11 &   27.24 &   0.14 &\hspace{1.0cm}3.17 &   &yes \\
 13  48  54.37  &  17  11  49.6 &19.13 &    1.90 &   0.99 &\hspace{1.0cm}3.60 &   &    \\
 13  49  18.52  &  35  24  15.7 &19.77 &   81.88 &   0.11 &\hspace{1.0cm}1.22 &   &yes \\
 14  06  35.67  &  62  25  43.3 &19.72 &   11.50 &   0.47 &\hspace{1.0cm}3.89&abs &yes \\
 14  34  13.05  &  16  28  52.7 &19.86 &    4.95 &   1.00 &\hspace{1.0cm}4.21&    &    \\
 14  53  29.01  &  48  17  24.9 &20.11 &    3.75 &   1.00 &\hspace{1.0cm}3.77&    &yes \\
 15  11  46.99  &  25  24  24.3 &19.95 &    1.39 &   1.00 &\hspace{1.9cm}3.719&BAL&yes \\
 15  20  28.14  &  18  35  56.1 &19.82 &    6.94 &   0.96 &\hspace{1.0cm}4.11&    &    \\
 15  24  24.35  &  07  31  29.9 &20.13 &    1.51 &   0.39 &\hspace{1.0cm}3.59&    &yes \\
 15  33  36.14  &  05  43  56.5 &19.84 &   28.29 &   0.99 &\hspace{1.0cm}3.93&    &    \\
 15  37  56.90  &  48  23  32.3 &20.00 &    3.07 &   0.97 &\hspace{1.0cm}1.34&    &yes \\
 16  37  08.29  &  09  14  24.6 &19.56 &    9.43 &   0.93 &\hspace{1.0cm}3.75&    &    \\
 17  02  53.54  &  23  57  58.0 &19.74 &   19.24 &   0.25 &              & unknown&yes \\
 17  20  02.17  &  24  55  48.8 &19.82 &   12.90 &   0.16 &\hspace{1.0cm}3.34&BAL &    \\
 22  28  14.39  & -08  55  25.7 &20.19 &    1.99 &   0.99 &\hspace{1.0cm}3.64&     &yes \\
 22  35  35.59  &  00  36  02.0 &20.14 &    4.32 &   0.98 &\hspace{1.0cm}3.87&     &yes \\
 23  50  22.39  & -09  51  44.3 &19.68 &    6.51 &   0.10 &\hspace{1.0cm}3.70&     &yes \\
\hline
\end{tabular}

The columns give: (1,2,3,4) similar to Table 2; (5) NN output (see
Section 4.2); (6) redshifts from NED (SDSS 072518.26+370518.3, SDSS
074711.15+273903.3 and SDSS 122027.96+261903.5 from Benn et al. 2002;
SDSS 122135.64+280613.8 from Mason et al. 2000), from this work or
from DR6; (7) BAL = broad absorption line QSO; abs = the Ly$\alpha$
line appears to be self-absorbed; (8) indicates the sources located
within the spectroscopic plates available in DR6.

 \end{minipage}
 \end{table*}
\subsection{The sample of high-$z$ QSO candidates}

In Section 3 we demonstrated the good performance of simple NNs for
separating, on the basis of optical photometry and radio data,
high-$z$ QSOs from the remaining classes in the labelled sample.
This evaluation was based on the outputs of $m=4248$ labelled
objects, using for each object a NN trained with the remaining 4247.
We adopted set B of input variables, which for threshold $y_c=0.1$
yielded completeness $96 \pm 14$ per cent and efficiency $62 \pm 9$
per cent. The $m$ NNs were applied to the unlabelled sample and the
resulting $m$ values of $y$ for each source, and their medians
$y_{\rm med}$, are shown in Fig. 6. The number of unlabelled sources
with median outputs exceeding $y_c=0.1$ is 58 (31 outside and 27
within the DR5 spectroscopic area), and these are our `high-$z$ QSO
candidates'. Their properties are listed in Table 5, including the
median of the NN outputs. The same parameter for the high-$z$ QSOs
in the labelled sample is listed in Table 2 (but in this case the
QSOs themselves were used in the training).

\subsection{Spectroscopic check of the QSO selection}

The quality of the selection of high-$z$ QSO candidates was tested
by comparison with spectroscopy from the NED database, from a
dedicated observing programme with ISIS at the William Herschel
Telescope, and from spectroscopic classifications from SDSS DR6. The
resulting classifications and redshifts are included in Table 5.

\subsubsection{Spectroscopic classifications from NED}

Counterparts of the 58 high-$z$ QSO candidates and the
4357 non-candidates (4415 unlabelled
sources) were sought in NED in February 2007, using a search radius of
5 arcsec.
A similar radius is used in the SDSS DR5 Quasar Catalog
(Schneider et al. 2007) to quote the association with a NED object.
Four of the candidates were spectroscopically
classified, all QSOs with $z >$ 3.3.
Benn et al. (2002) identified three of them, with redshifts 4.33, 4.17
and 3.694, the remaining candidate, with $z=3.305$, was identified
by Mason et al. (2000).

Amongst the non-candidates, 382 were associated with QSOs
(blazars excluded) with secure redshifts, and none of them had $z
\ge 3.6$.  Another 13 QSOs had ambiguous or uncertain redshifts but
none were consistent with $z \ge 3.6$.

\begin{figure*}
\par
\centerline{\includegraphics[width=140mm]{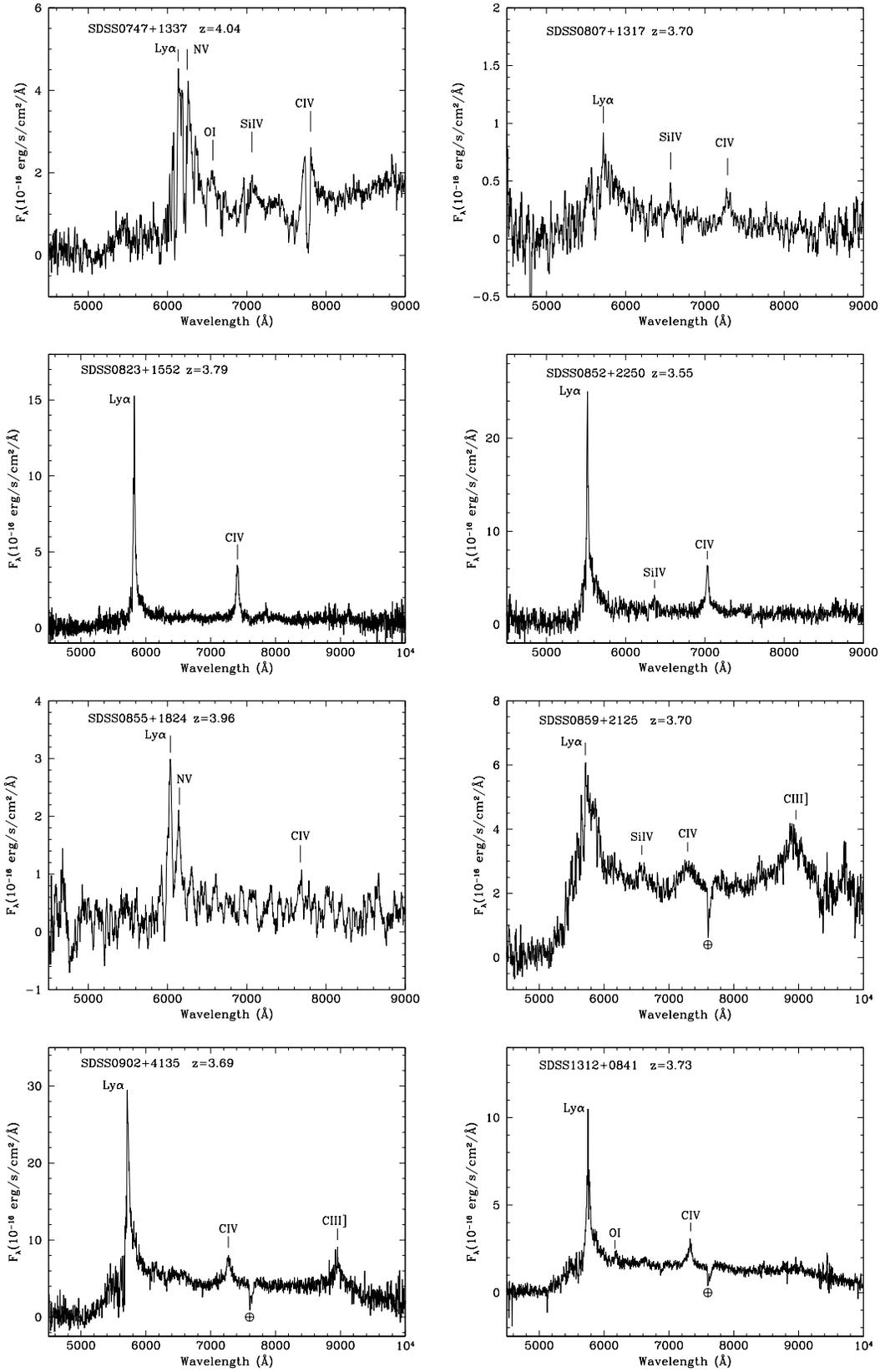} }
\par
\caption{WHT spectra of 21 of the 58 NN $y_{\rm med} \ge $ 0.1
candidates (six also included in DR6), including 17 $z \ge 3.6$
QSOs, and four QSOs at $z=3.34$, $z=3.42$, $z=3.55$ and $z=3.59$.
Emission features are labelled by ion. Symbol $\oplus$ indicates the
position of the atmospheric absorption band O$_2$ A.}
\end{figure*}

\setcounter{figure}{6}
\begin{figure*}
\par
\centerline{\includegraphics[width=140mm]{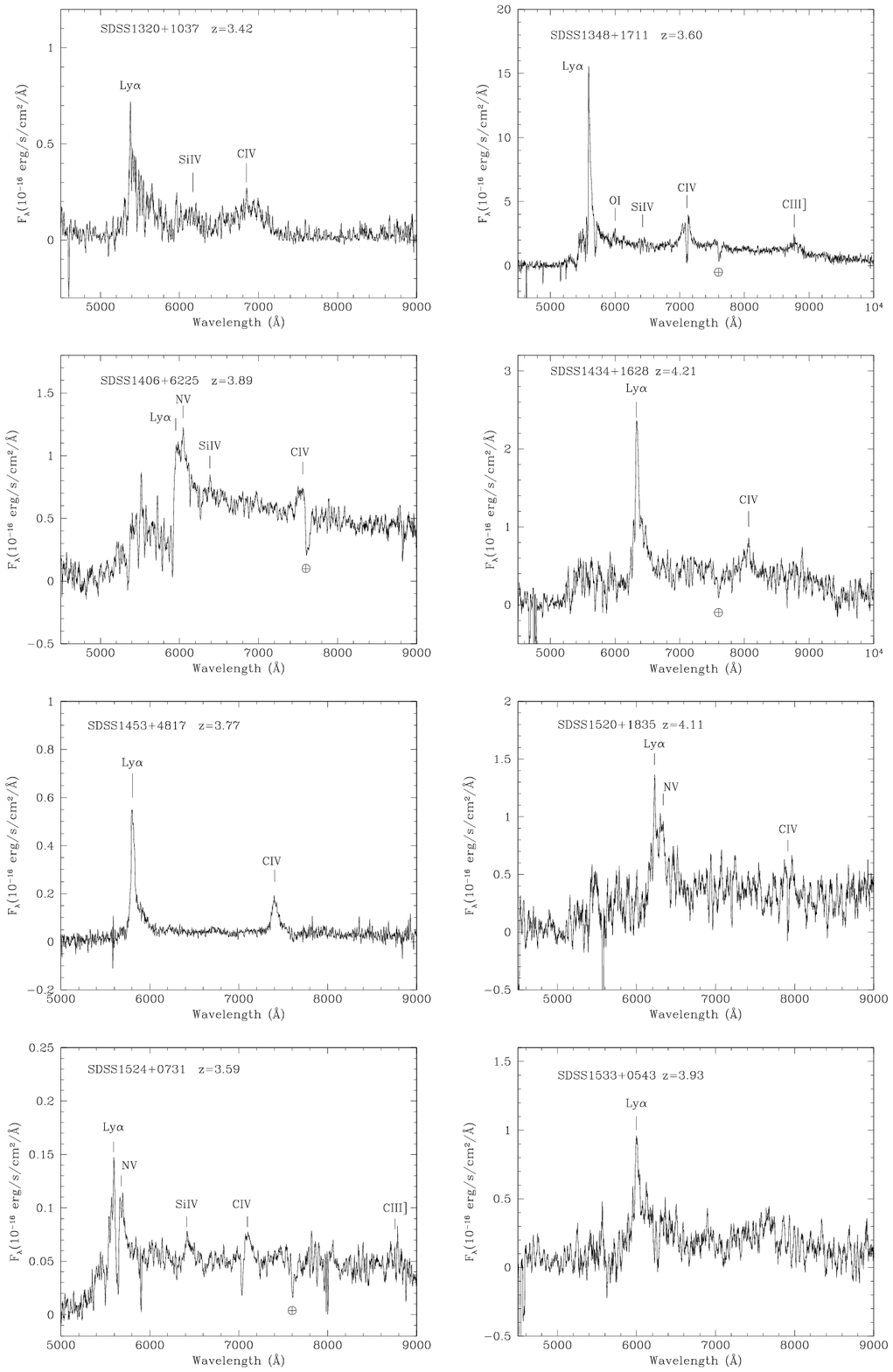} }
\par
\caption{Continued}
\end{figure*}

\setcounter{figure}{6}
\begin{figure*}
\par
\centerline{\includegraphics[width=140mm]{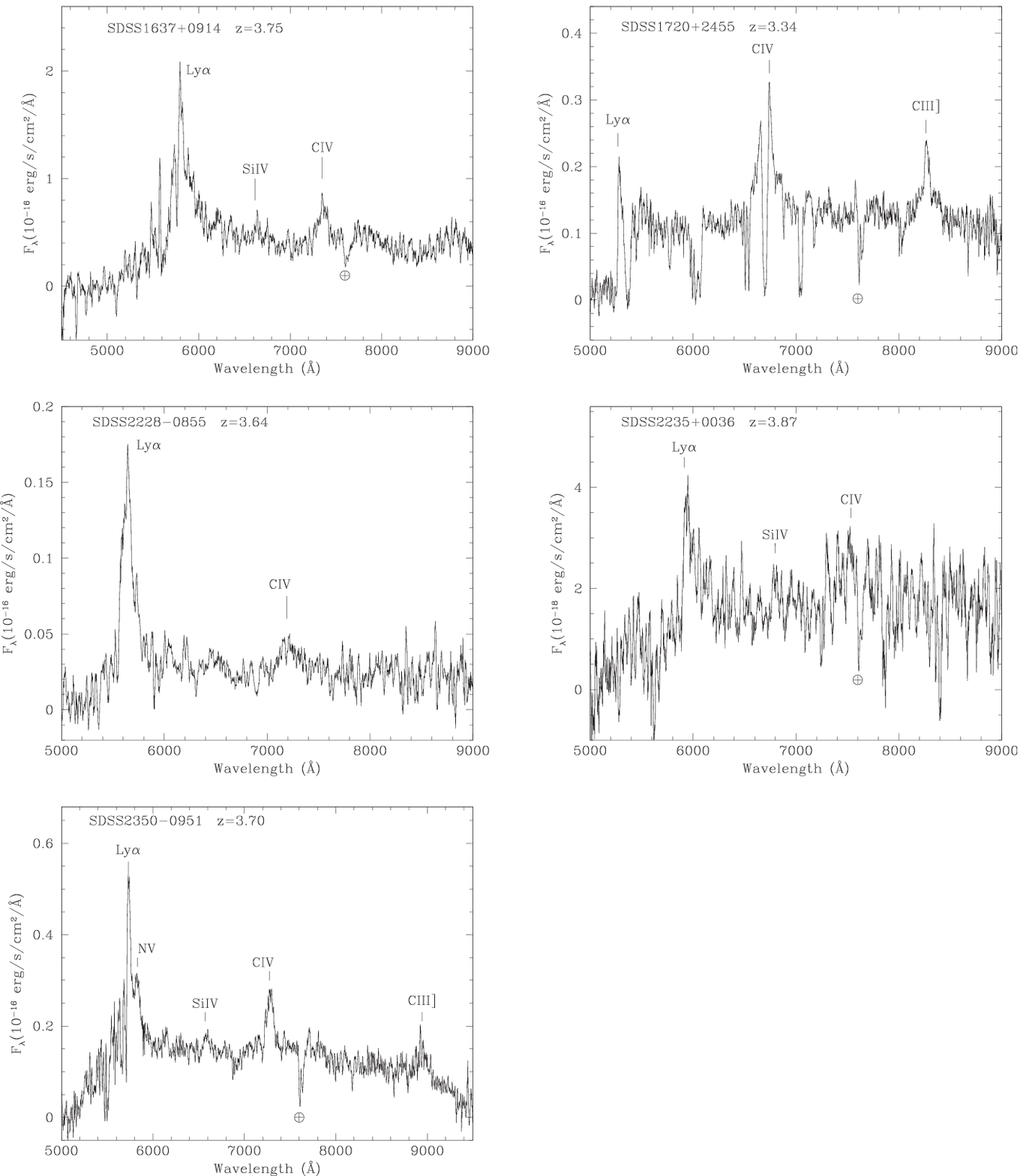}}
\par
\caption{Continued}
\end{figure*}

\subsubsection{Spectroscopy with ISIS}

Spectra of 27 of the remaining 54 candidates were obtained with the
WHT's ISIS dual-arm spectrograph in two runs on 2007 April 3, 4, 6
and 7 and July 8, 9 and 10. The R158R grating was used on the red
arm, yielding a wavelength range 5300$-$10000 \AA\ with dispersion
1.8 \AA\ pixel$^{-1}$. On the blue arm the R300B grating was used,
giving a spectral range 3000$-$6000 \AA\ with dispersion 0.9 \AA\
pixel$^{-1}$. Exposure times were 600 s. Spectrophotometric standard
stars were observed in order to calibrate the instrumental spectral
response. Seeing was typically better than 1 arcsec and the slit
width was set to 1 arcsec, yielding spectral resolution, as
measured from sky lines, of 7.7 \AA\ and 4.5 \AA\ in the red and
blue arms respectively. Standard data reduction was carried out
using the IRAF\footnote{IRAF is distributed by the National Optical
Astronomy Observatories, which is operated by Association of
Universities for Research in Astronomy, Inc., under cooperative
agreement with the National Science Fundation} package. Arc-lamp
exposures were used for wavelength calibration, giving solutions
with residuals $< 0.1$ \AA.

We observed 15 sources in April, prioritising the candidates with
higher NN output $y$, brighter $r$ magnitude, and at lower air mass,
regardless of their location inside or outside the DR5 spectroscopic
area. All 15 sources were classified as QSOs and their redshifts
were determined as the average of the values measured from
individual emission-line centroids, usually Ly$\alpha$, Si {\sc iv},
and C {\sc iv}. 13 of the QSOs have $3.60 \le z \le 4.21$, and the
remaining two $z=3.55$ and $z=3.42$.

The 12 sources observed in July consist of {\it all} the remaining
candidates with right ascension in the range 13 -- 24 hours and
without spectra in DR6. The results of the spectroscopic
classification were as follows. 11 sources were classified as QSOs:
four with $3.6 \le z \le 3.9$, four with $1.07 \le z \le 1.34$ and
three with $z=3.17$, $z=3.34$ and $z=3.59$. One candidate remained
unclassified due to both low signal-to-noise and lack of clear
absorption or emission features in the spectrum.

Fig. 7 shows the spectra of the 21 candidates identified with ISIS
as $z \ge 3.2$ QSOs (17 of them with $z \ge 3.6$).

\subsubsection{Spectroscopy from SDSS DR6}

SDSS DR6 (CAS {\tt SpecObj} view) includes spectra of nine of the
remaining 27 candidates (and also of six of those already observed
with ISIS in April, and two from the NED).  These nine include four
QSOs with $3.6 \le z \le 3.8$, one at $z=3.40$, two galaxies with
$z=0.45$ and $z=0.58$, and two sources labelled as unknown.

Amongst the 4357 non-candidates, 898 have spectra from DR6,
and are classified as stars or galaxies (71), unknown (62) or QSOs (765).
The latter include some of the QSOs with spectra from NED cited in
Section 4.3.1.
%Two QSOs have $z \ge 3.6$, but one of the SDSS redshifts
%is incorrect, the other is doubtful.
Two QSOs have $z \ge 3.6$, but their SDSS redshifts are incorrect (see the
DR6 spectra in Fig. 8).
For SDSS 075558.88+113210.9, with quoted $z=4.340$ (confidence=0.51), we measured from the emission lines of Mg {\sc ii}
(erroneously taken as Ly$\alpha$ by the SDSS pipelines) and C {\sc
iii]} a redshift $z=1.295$. The spectrum of SDSS 161836.09+153313.5,
with $z=4.376$ (confidence=0.00), resembles that found for
sources SDSS 130941.36+112540.1 and SDSS 153420.23+413007.5
(see Section 2), classified in DR5 as QSOs with redshifts $z=4.395$
and $z=4.426$, but with revised values in DR5Q
$z=1.362$ and $z=1.400$, due to the re-interpretation of the
assumed Ly$\alpha$ emission line as Mg
{\sc ii}. A similar revision gives $z = 1.322$ for SDSS 161836.09+153313.5.

\begin{figure*}
\par
\centerline{\includegraphics[width=140mm]{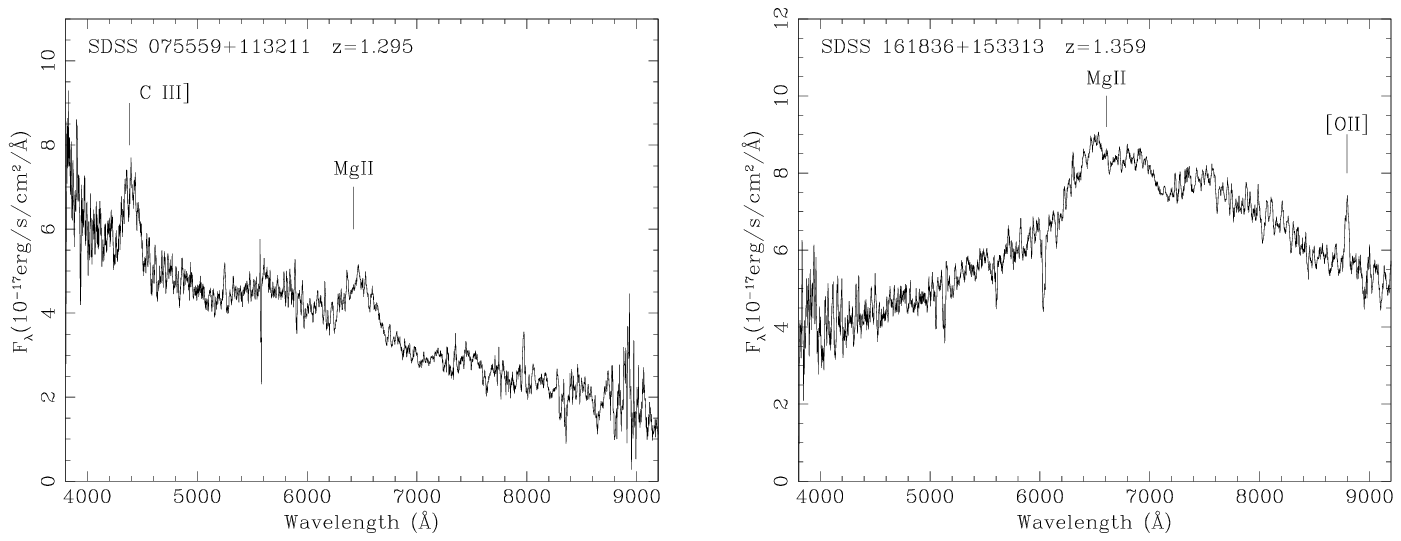} }
\par
\caption{DR6 spectra of two non-candidates with revised redshifts
$z=1.29$ and $z=1.32$ listed in SDSS-DR6 as $z \ge 3.6$ QSOs, due to
misidentification between the Ly$\alpha$ and Mg {\sc ii} emission
lines.}
\end{figure*}

\subsubsection{Performance of the selection of high-redshift QSOs}

Spectra are available for 40 candidates, obtained from
NED, SDSS DR6, or observed for this work,
and 24 of them are confirmed $z \ge 3.6$ QSOs.

Figure 9 shows a plot of the NN output $y_{\rm med}$ versus
$r$ magnitude for the 58 high-$z$ QSO candidates.
Different symbols correspond to the different spectral classes
(high-$z$ QSOs are shown as diamonds).
The two panels separate the candidates located within and outside the
spectroscopic area available in DR6 (as noted in the last column of
Table 5).
Fig. 9 shows a larger fraction of high-$z$ QSOs among
the candidates with higher NN outputs, a trend also evident in Fig. 2.
The efficiency for $0.55 \le y \le 1$ is 91 per cent (20 $z \ge 3.6$ QSOs
out of 22 candidates with available spectra), dropping to
22 per cent (four of 18) for $0.1 \le y < 0.55$. Spectra are available
for all 20 high-$z$ QSO candidates with $13^{\rm h} <$ RA $<  24^{\rm h}$,
and this set therefore forms a complete subsample with regard to the
distribution of NN outputs. With 12 confirmed $z \ge 3.6$ QSOs out of 20
candidates, the efficiency from this sample is $60 \pm 17$ per cent.
This value is in good agreement with the efficiency 24/40 = $60 \pm 12$
per cent for  the total sample of candidates with spectra, and with the
$62 \pm 9$ per cent efficiency obtained within the labelled sample (Section 3).

\begin{figure*}
\includegraphics[width=179mm]{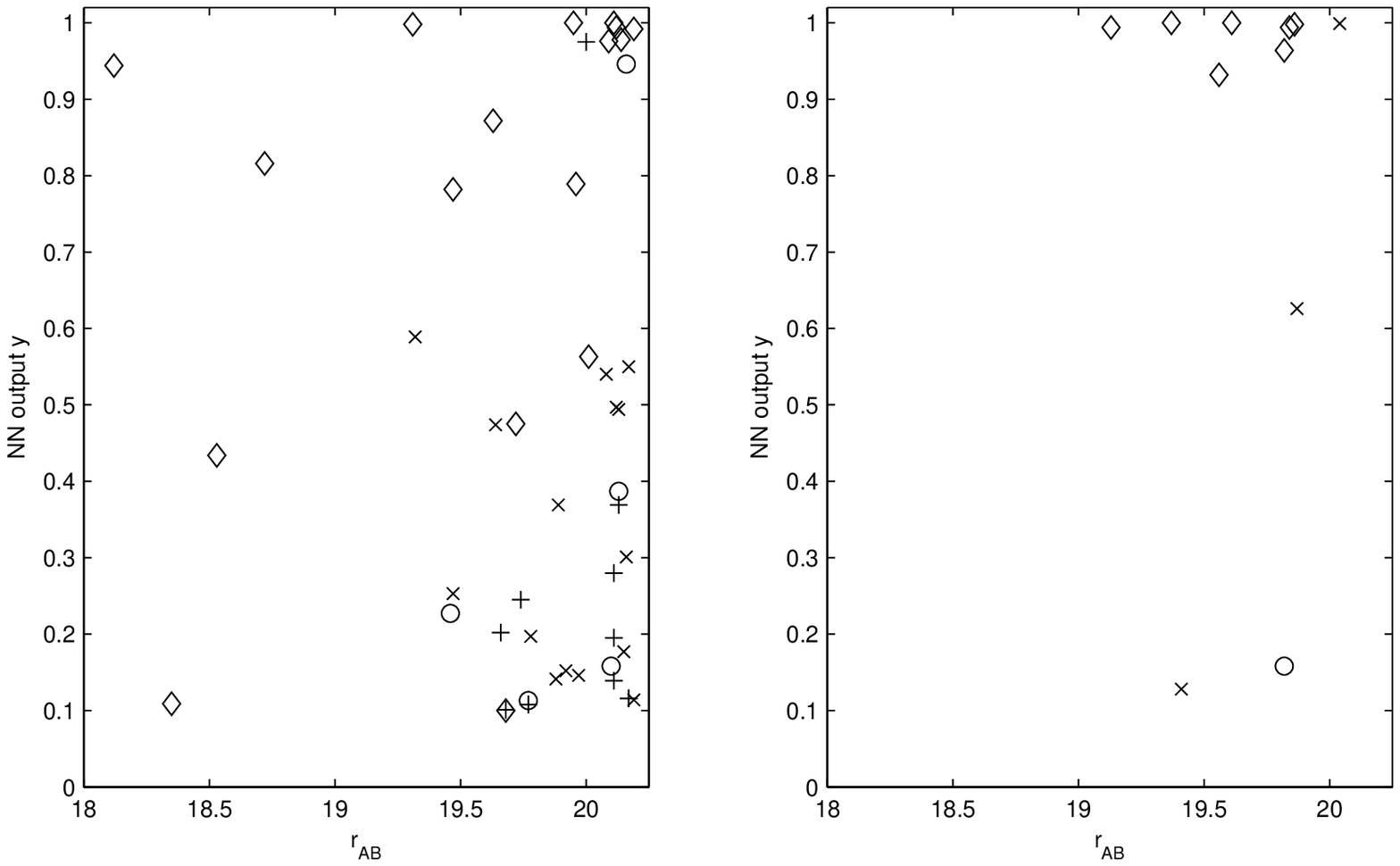} \caption{ NN output
$y$ versus $r$ magnitude for the 58 high-$z$ QSO candidates, located
either within ({\it left panel}) or outside ({\it right panel}) the
spectroscopic area available in DR6. $\diamond$: QSOs with $z \ge 3.6$,
$\circ$: QSOs with $3.2 \le z < 3.6$, +: other classification or
unknown, $\times$: without spectrum.}
\end{figure*}

None of the non-candidates in the unlabelled sample with available
spectrum from NED or DR6 is a $z \ge 3.6$ QSO, therefore we have no
evidence of incompleteness, with respect to DR5 unlabelled
sources with matches in any of these databases. In fact a
good completeness was expected for the matches with the DR6
spectroscopic survey, since the selection of sources,
classification and measured properties respond to the same
scheme as for the DR5 spectroscopic catalogue used for the training,
where we had found 96 per cent completeness. However,
the NED database provides spectroscopic identifications assigned
from other surveys, and the absence of NED high-$z$ QSOs among the
non-candidates gives therefore independent evidence that the
classifier has a good completeness in its extension
to DR5 unlabelled sources.
Three DR5 unlabelled sources are identified as $z \ge 3.6$ QSOs in NED,
all of them selected as high-$z$ QSO candidates in our work.

%We note (see Sect. 4.3.1) that
%four high-$z$ candidates had NED identifications, of which
%three were $z \ge 3.6$ QSOs and the other a QSO with $z=3.3$, and 395
%non-candidates were identified with QSOs, all with $z < 3.6$.

\section{Discussion and conclusions}

In this paper we aimed to obtain a sample of $z \ge 3.6$ QSOs,
starting from an initial complete sample of 8665 FIRST sources with
star-like counterparts in the SDSS DR5 photometric survey, of which
4250 have spectra in DR5, 52 of them being $z \ge 3.6$ QSOs. We
found that simple supervised NNs, trained on the sources with DR5
spectra, and using optical photometry and radio data as input
parameters, allow separation of high-$z$ QSOs from the remaining
spectral classes with 96 per cent completeness and 62 per cent
efficiency. The application of the NNs to the sample of
4415 sources without DR5 spectra yielded 58 high-$z$ QSO candidates.

We obtained spectra of 27 of the 58 candidates, and 17 were confirmed
as high-$z$ QSOs. Spectra of 13 additional candidates from the NED and
from DR6 revealed seven more high-$z$ QSOs,
%[three identified from Benn et al. (2002) and four from DR6]
yielding a total 40 candidates with spectra available, of which 24 are
high-$z$ QSOs. The number of high-$z$ QSOs was then increased from
52 in the initial sample to 76 (a factor 1.46).

The overall efficiency in the selection of new high-$z$ QSOs is
$60 \pm 12$ per cent (24/40). The estimate from a subsample
unbiased with respect to the NN outputs is $60 \pm 17$ per cent
(12/20), and both values are in good agreement with the $62 \pm 9$
per cent efficiency obtained for the DR5 labelled sample (Section 3).

None of the non-candidates with spectra available from NED or DR6 is a
$z \ge 3.6$ QSO, therefore we have no evidence of incompleteness regarding
high-$z$ QSOs with matches in these catalogues. Since the NED
spectroscopic identifications are assigned from a variety of surveys different
than SDSS, the database provides a blind test of the good completeness of the
classifier for DR5 unlabelled sources.

The efficiencies found are well above the values obtained for
previous samples of RL high-$z$ QSOs, based on less accurate optical
photometry and with fewer wavelength bands than SDSS, although
already highly complete ($\ge 95$ per cent) regarding optical colour
selection. The efficiencies for various samples,
summarised by Carballo et al. (2006)
for $z \ge 3.7$, are $12-13$ per cent (Holt et al. 2004, Carballo et
al. 2006; $S_{\rm 1.4 ~GHz}>$ 1 mJy, APM POSS $E$, $O$), 6 per cent
(Hook et al. 2002; $S_{\rm 5 ~GHz} >$ 30 mJy and radio flat, APM
POSS $E$, $O$), and 19 per cent (Snellen et al. 2001; $S_{\rm 5
~GHz} >$ 30 mJy and radio flat, APM UKST $B$, $R$, $I$).

Adopting for the 18 candidates which still lack spectroscopy
a weighted efficiency of 37 per cent (four
candidates with $y \ge 0.55$ and 14 with $y < 0.55$,
with expected efficiencies of 91 and 22 per cent respectively),
we calculate $\sim$ 7 additional $z \ge 3.6$ QSOs.
The FIRST-DR5 sample of high-$z$ QSOs is thus
expected to contain $\sim 83$ QSOs (52+24+7).
Adopting as a lower limit for completeness the nominal value of
96 per cent found for the labelled sample, we calculate for the set of
31 high-$z$ QSOs obtained by the classifier (24 discovered and
7 predicted) a minimum $ \sim 1$ missed high-$z$ QSO.

The NNs found 31 contaminants,  i.e. non high-$z$ QSOs
erroneously selected as high-$z$ QSOs, in the labelled sample,
with a fraction $61 \pm 14$ per cent (19/31) being QSOs
with $3.2 \le z \le 3.6$. Among the 40 high-$z$
QSO candidates with available spectra we found 16 contaminants,
seven of them being QSOs with $3.15 \le z \le 3.6$, confirming a
high rate of QSOs with $z$ near the threshold $z=3.6$ among the
contaminants, 7/16 = $44 \pm 17$ per cent.

Our results allow us to obtain an estimate of the incompleteness of
SDSS for the spectroscopic classification of FIRST high-$z$ QSOs.
47 of the high-$z$ QSO candidates are located in the spectroscopic
area covered by DR6 (Table 5, Fig. 9 left panel), and 17 of them are
$z \ge 3.6$ QSOs, ten included in the DR6 spectroscopic catalogue
and seven not included. 15 candidates in this area still lack spectroscopy,
and assuming for them a weighted efficiency of 31 per cent
(two candidates with $y \ge 0.55$ and 13 with $y < 0.55$), we expect
another four high-$z$ QSOs. From this calculation we estimate $11$
FIRST high-$z$ QSOs missed by SDSS (7 QSOs and $4$ candidates), which
when compared to 62 (52+10) identifications yields an incompleteness
of SDSS for the spectroscopic classification of
FIRST $3.6 \le z \le 4.6$ QSOs of $\sim$ 15 per cent (11/73) for
$r \le 20.2$.

The definition of the original sample of 52 high-$z$ FIRST QSOs excluded
lobe-dominated morphologies and ``narrow-lined'' QSOs, and included
QSOs with BALs or self-absorbed at Ly$\alpha$. The same conditions hold
for the larger sample of 76 QSOs, obtained from the application of the NNs
trained using these 52 objects to DR5 photometric sources without
spectra in DR5, and covering a slightly wider region of input space than that
used by SDSS for QSO targetting.

In a future paper we plan to analyze the optical luminosity function of
FIRST-SDSS QSOs at $3.6 \le z \le 4.6$ on the basis of this sample.
Concurrently we expect to carry out spectroscopic observations of the
18 candidates without spectra. Given the efficacy of our approach,
we intend to extend the sample using more updated SDSS data releases,
increasing the number of training sources and the number of high-$z$ QSO
candidates, for which subsequent spectroscopy will be planned. We envisage
using additional infrared data via UKIDSS (UKIRT Infrared Deep Sky Survey).

%Various sources located within the DR6 spectroscopic area but
%unlabelled in DR6 were
%classified by the NN as high-$z$ candidates, and confirmed as such
%in this work, showing that the statistical model provides a {\it
%generalisation} of the data.
%We used the FIRST-DR5 labelled sources as examples from which a
%mathematical model was built, tracing the best linear separation
%between high-$z$ QSOs and the remaining classes in the parameter
%space defined by SDSS photometry and radio data.

\section*{Acknowledgments}

We are grateful to the referee for a prompt and useful report, which
improved the paper. RC, JIGS, CRB, and FJL acknowledge financial support
from the Spanish Ministerio de Educaci\'on y Ciencia under project
AYA 2005-00055.  We acknowledge the Isaac Newton Group's service programme
for a generous allocation of observing time for this project and we thank
ING staff for carrying out the observations.

Funding for the creation and distribution of the SDSS Archive has been
provided by the Alfred P. Sloan Foundation, the Participating
Institutions, the National Aeronautics and Space Administration, the
National Science Foundation, the U.S. Department of Energy, the
Japanese Monbukagakusho, and the Max Planck Society. The SDSS Web site
is http://www.sdss.org/. The SDSS is managed by the Astrophysical
Research Consortium (ARC) for the Participating Institutions. The
Participating Institutions are The University of Chicago, Fermilab,
the Institute for Advanced Study, the Japan Participation Group, The
Johns Hopkins University, the Korean Scientist Group, Los Alamos
National Laboratory, the Max-Planck-Institute for Astronomy (MPIA),
the Max-Planck-Institute for Astrophysics (MPA), New Mexico State
University, University of Pittsburgh, University of Portsmouth,
Princeton University, the United States Naval Observatory, and the
University of Washington. This research has made use of the NASA/IPAC
Extragalactic Database (NED) which is operated by the Jet Propulsion
Laboratory, California Institute of Technology, under contract with
the National Aeronautics and Space Administration.


\begin{thebibliography}{99}

\bibitem{} Adelman-McCarthy J. et al., 2007, ApJS 172, 634
\bibitem{} Bailer-Jones C.A.L., Irwin M., von Hippel T., 1998, MNRAS, 298,
361
\bibitem{} Ball N.M., Loveday J., Fukugita M., Nakamura O.,
Okamura S., Brinkmann J., Brunner R.J., 2004, MNRAS, 348, 1038
\bibitem{} Ball N.M., Brunner R.J., Myers A.D., Tcheng D., 2006, ApJ, 650, 497
\bibitem{} Ball N.M., Brunner R.J., Myers A.D., Strand N.E.,
Alberts S.L., Tcheng D., Llor\`a X., 2007, ApJ, 663, 774
\bibitem{} Ball N.M., Brunner R.J., Myers A.D., Strand N.E.,
Alberts S.L., Tcheng D., 2008, ApJ in press, astro-ph 08043413
\bibitem{} Bazell D., Miller D.J., SubbaRao M., 2006, ApJ 649, 678
\bibitem{} Becker R.H., White R.L., Helfand D.J., 1995, ApJ, 450, 559
\bibitem{} Benn C.R., Vigotti M., Pedani M., Holt J., Mack K.-H., Curran
R., S\'anchez S.F., 2002, MNRAS, 329, 221
\bibitem{} Bertin E., Arnouts S., 1996, A\&AS, 117, 393
\bibitem{} Bishop C.M., 1995, Neural Networks for Pattern Recognition,
Oxford University Press
\bibitem{} Carballo R., Cofi\~no A.S., Gonz\'alez-Serrano J.I., 2004, MNRAS, 353, 211
\bibitem{} Carballo R., Gonz\'alez-Serrano J.I., Montenegro-Montes F.M.,
Benn C.R., Mack K.-H., Pedani M., Vigotti M., 2006, MNRAS, 370, 1034
\bibitem{} Cirasuolo M., Magliocchetti M., Gentile G., Celotti A., Cristiani
S., Danese L., 2006, MNRAS, 371, 695
\bibitem{} Claeskens J.-F., Smette A., Vandenbulcke L., Surdej J., 2006,
MNRAS, 367, 879
\bibitem{} Collister A.A., Lahav O., 2004, PASP, 116, 345
\bibitem{} Crom S.M., Smith R.J., Boyle B.J., Shanks T., Miller L.,
Outram P.J. Loaring N.S., 2004, MNRAS, 349, 1397
\bibitem{} Cutri R.M. et al., 2003, The IRSA 2MASS All-Sky Point Source
Catalog. NASA/IPAC Infrared Science Archive
\bibitem{} de Vries W.H., Becker R.H., White R.L., 2006, AJ 131, 666
\bibitem{} Firth A.E., Lahav O., Somerville R.S., 2003, MNRAS, 339, 1195
\bibitem{} Folkes S.R., Lahav O., Maddox S.J., 1996, MNRAS, 283, 651
\bibitem{} Gao D., Zhang Y.-X., Zhao Y.-H., 2008, MNRAS, 386, 1417
\bibitem{} Hagan M.T., Menhaj M., 1994, IEEE Trans. Neural Networks, 5, 989
\bibitem{} Holt J., Benn C.R., Vigotti M., Pedani M.,  Carballo R.,
Gonz\'alez-Serrano J.I., Mack K.-H., Garc\'\i a B., 2004, MNRAS,
348, 857
\bibitem{} Hook I.M., McMahon R.G., Shaver P.A., Snellen I.A.G., 2002, A\&A,
391, 509
\bibitem{} Ivezic Z. et al. 2004, in G.T. Richards \& P.B. Hall, eds.,
AGN Physics with the Sloan Digital Sky Survey. ASP Conf. Ser. 311, 347.
San Francisco
\bibitem{} Jiang L., Fan X., Ivezic Z., Richards G.T., Schneider D.P., Strauss
M.A., Kelly B.C., 2007, ApJ, 656, 680
\bibitem{} Kormendy J., Richstone D., 1995, ARA\&A, 33, 581
\bibitem{} Lahav O., Naim A., Sodr\'e L. Jr, Storrie-Lombardi M.C., 1996,
MNRAS, 283, 207
\bibitem{} Magorrian J. et al., 1998, AJ, 115, 2285
\bibitem{} Mason K.O. et al., 2000, MNRAS, 311, 456
\bibitem{} Richards G.T. et al., 2002, AJ, 123, 2945
\bibitem{} Richards G.T. et al., 2004, ApJSS, 155, 257
\bibitem{} Richards G.T. et al., 2006, AJ, 131, 2766
\bibitem{} Rohde D.J., Drinkwater M.J., Gallagher M.R., Downs T., Doyle M.T.,
2005, MNRAS, 360, 69
\bibitem{} Schlegel D.J., Finkbeiner D.P., Davis M., 1998, ApJ, 500, 525
\bibitem{} Schneider D.P. et al., 2007, AJ, 134, 102
\bibitem{} Snellen I.A.G., McMahon R.G., Dennett-Thorpe J., Jackson N.,
Mack K.-H., Xanthopoulos E., 2001, MNRAS, 325, 1167
\bibitem{} Suchkov A. A., Hanisch R.J., Margon B., 2005, AJ 130, 2439
\bibitem{} Vigotti M., Carballo R., Benn C.R., de Zotti G., Fanti
R., Gonz\'alez-Serrano J.I., Mack K.-H., Holt J., 2003, ApJ, 591, 43
\bibitem{} Voges W. et al., 1999, A\&A, 349, 389
\bibitem{} Voges W. et al., 2000, IAU Circ., 7432, 3
\bibitem{} Weinstein M.A. et al., 2004, ApJSS, 155, 243
\bibitem{} York D.G. et al., 2000, AJ, 120, 1579

\end{thebibliography}
\end{document}